\documentclass[11pt]{article}
\pdfoutput=1
\usepackage{jheppub}
\usepackage{tabu}
\usepackage[vcentermath]{youngtab}
\usepackage[usenames,dvipsnames,table]{xcolor}
\usepackage{graphicx,amsmath,amssymb,amsthm,multirow,array,bm,bbm,esint}
\usepackage[mathscr]{eucal}
\usepackage[bbgreekl]{mathbbol}
\usepackage{epsf,amsfonts}
\usepackage{slashed}
\usepackage[numbers,sort&compress]{natbib}
\usepackage{minibox}
\usepackage{rotating}
\usepackage{pdflscape}
\usepackage{array,tikz-cd}
\usepackage{subcaption}
\usepackage{framed}
\usepackage{mathtools}
\usepackage{tikz} 
\usepackage[compat=1.1.0]{tikz-feynman}
\usepackage{feynmf}
\usepackage{float}
\usepackage{verbatim}

\usepackage{calligra}

\usepackage[normalem]{ulem}

\usepackage{slashed}

\newcommand{\be}{\begin{equation}}
\newcommand{\ee}{\end{equation}}
\newcommand{\bea}{\begin{eqnarray}}
\newcommand{\eea}{\end{eqnarray}}

\newcommand{\mt}[1]{\textrm{\tiny #1}}

\DeclareMathAlphabet{\mathcalligra}{T1}{calligra}{m}{n}
\DeclareFontShape{T1}{calligra}{m}{n}{<->s*[2.2]callig15}{}
\newcommand{\scriptr}{\mathcalligra{r}\,}

\makeatletter
\newcommand{\doubletilde}[1]{{%
  \mathpalette\double@tilde{#1}%
}}
\newcommand{\double@tilde}[2]{%
  \sbox\z@{$\m@th#1\tilde{#2}$}%
  \ht\z@=.9\ht\z@
  \tilde{\box\z@}%
}
\makeatother

\title{Scrambling in Hyperbolic Black Holes: shock waves and pole-skipping}

\author[]{Yongjun Ahn,}
\author[]{Viktor Jahnke,}
\author[]{Hyun-Sik Jeong,}
\author[]{and  Keun-Young Kim}

\affiliation[]{School of Physics and Chemistry, Gwangju Institute of Science and Technology, 123 Cheomdan-gwagiro, Gwangju 61005, Korea}

\emailAdd{yongjunahn619@gmail.com}
\emailAdd{viktorjahnke@gist.ac.kr}
\emailAdd{hyunsik@gist.ac.kr}
\emailAdd{fortoe@gist.ac.kr}

\abstract{ We study the scrambling properties of $(d+1)$-dimensional hyperbolic black holes. Using the eikonal approximation, we calculate out-of-time-order correlators (OTOCs) for a Rindler-AdS geometry with AdS radius $\ell$, which  is dual to a $d-$dimensional conformal field theory (CFT) in hyperbolic space with temperature $T = 1/(2 \pi \ell)$. We find agreement between our results for OTOCs and previously reported CFT calculations. 
For more generic hyperbolic black holes, we compute the butterfly velocity in two different ways, namely: from shock waves and from a pole-skipping analysis, finding perfect agreement between the two methods. The butterfly velocity $v_B(T)$  nicely interpolates between the Rindler-AdS result $v_B(T=\frac{1}{2\pi \ell})=\frac{1}{d-1}$ and the planar result $v_B(T \gg \frac{1}{\ell})=\sqrt{\frac{d}{2(d-1)}}$\,.
}

\begin{document}
\maketitle

%\tableofcontents

\section{Introduction} \label{sec-1}

In recent years, out-of-time-order correlators (OTOCs)
\begin{equation}
F(t, {\bf x}) := \langle V_0(0) W_{\bf x}(t) V_0(0) W_{\bf x}(t) \rangle \,,
\end{equation}
have been recognized as very useful tools to diagnose many-body quantum chaos\footnote{See~\cite{Cotler:2017jue,deMelloKoch:2019rxr,Murthy:2019fgs,Ma:2019ocx,Nosaka:2018iat} for studies connecting/comparing OTOCs with other notions of quantum chaos.}. Here,  $V$ and $W$ are general local operators and we denote their spatial dependence as subscripts, i.e., $W_{\bf x}(t)$.  In the case of holographic theories, OTOCs have a dual gravitational description in terms of a high-energy collision that takes place close to the black hole horizon~\cite{BHchaos1,BHchaos2,BHchaos3,BHchaos4}. This leads to a  simple and universal result 
\be
\frac{\langle V_0(0) W_{\bf x}(t) V_0(0) W_{\bf x}(t) \rangle}{\langle V_0(0) V_0(0)\rangle \langle W_{\bf x}(t) W_{\bf x}(t) \rangle} = 1-\varepsilon_{\Delta_V \Delta_W} e^{ \lambda_L\left(t-t_*-\frac{|{\bf x}|}{v_B} \right) } \quad \text{for} \quad t_d << t \lesssim t_*\,,
\ee
where $\lambda_L$ is the Lyapunov exponent, $v_B$ is the butterfly velocity, and the $t_*$ is the scrambling time.  All these parameters are determined from the geometry near the black hole horizon, and they are universal in the sense that they do not depend on the operators $V$ and $W$. The prefactor $\varepsilon_{\Delta_V \Delta_W}$ is a non-universal piece that contains information about the operators in the OTOC. The dissipation time $t_d$ controls the decay of two-point functions, i.e., $\langle V(0)V(t) \rangle \sim e^{-t/t_d}$.

Despite the existence of a very extensive  literature about the holographic description of chaos\footnote{See, for instance, the recent reviews~\cite{Sarosi:2017ykf,Jahnke:2018off}.}, it is very difficult to find examples where OTOCs can be calculated in both sides of the AdS/CFT duality \cite{duality1,duality2,duality3}. The only cases where calculations were done in both sides are: BTZ black holes/2-dimensional CFTs~\cite{BHchaos4,Roberts:2014ifa,Poojary:2018esz,Jahnke:2019gxr,Cotler:2018zff,Haehl:2018izb}, and AdS$_2$ gravity/SYK-like models~\cite{Kitaev-2014,Maldacena:2016hyu,Jensen:2016pah,Maldacena:2016upp,Engelsoy:2016xyb}. In higher dimensional cases, there are some OTOC results for CFTs in hyperbolic space~\cite{Perlmutter:2016pkf}, which, however, have not yet been reproduced by holographic calculations.

In this work, we fill this gap. We calculate OTOCs for an AdS-Rindler geometry in $(d+1)-$dimensions for $d \geqslant 2 $. This geometry is dual to a $d-$dimensional CFT in hyperbolic space. We find agreement between our holographic calculations and the previously reported CFT results~\cite{Perlmutter:2016pkf}. For more generic black holes, we compute the butterfly velocity in two different ways, namely: from shock waves and from a pole-skipping analysis, finding perfect agreement between these two methods. The butterfly velocity $v_B(T)$ nicely interpolates between the AdS-Rindler result $v_B\left(T=\frac{1}{2\pi \ell}\right)=\frac{1}{d-1}$ and the planar result $v_B(T \gg \frac{1}{\ell})=\sqrt{\frac{d}{2(d-1)}}$\,.

This paper is organized as follows. In section \ref{sec-hypBHs}, we briefly review the geometry of hyperbolic black holes in AdS spacetime, and  discuss the hyperbolic slicing of AdS forming the Rindler wedge. In section \ref{sec-shock}, we use the eikonal approximation to derive OTOCs from bulk shock wave collisions. In section \ref{sec-poleskip}, we obtain the Lyapunov exponent and the butterfly velocity using a pole-skipping analysis. We discuss our results in section \ref{sec-disc}. We relegate some technical details to Appendix \ref{AppA}.

\section{Hyperbolic black holes in AdS spacetime}\label{sec-hypBHs}

\subsection{General hyperbolic black holes}

We consider the $(d+1)-$dimensional Einstein-Hilbert action 
\begin{equation}
S= \frac{1}{16\pi G_N} \int d^{d+1}x \sqrt{-g} \left( R+\frac{d(d-1)}{\ell^2}\right)\,,
\end{equation}
and, as a classical solution, the hyperbolic black holes of the form
\be
ds^2=-f(r)dt^2+\frac{dr^2}{f(r)}+r^2 dH_{d-1}^2\,,
\label{eq-metricS}
\ee
with the emblackening factor 
\be
f(r)=\frac{r^2}{\ell^2}-1-\frac{r_0^{d-2}}{r^{d-2}}\left( \frac{r_0^2}{\ell^2}-1\right)\,.
\label{eq-f}
\ee
Here, $\ell$ denotes the AdS length scale and $dH_{d-1}^2=d\chi^2+\sinh^2\chi d\Omega_{d-2}^2$ is the line element (squared) of the $(d-1)-$dimensional hyperbolic space $H_{d-1}$. ($d\Omega_{d-2}$ is the line element of a unit sphere $S^{d-2}$.) The horizon is located at $r=r_0$, while the boundary is located at $r=\infty$. 

These coordinates only cover the exterior region ($r \geq r_0$) of the black hole. The maximally extended spacetime (the two-sided eternal black hole geometry) can be described by introducing the Kruskal-Szekeres coordinates $U,V$ as
\begin{equation}
\begin{split}
U&=+e^{\frac{2\pi}{\beta}\left(r_*-t\right)}\,,\,\,V=-e^{\frac{2\pi}{\beta}\left(r_*+t\right)} \,\,\,(\text{left exterior region})  \\
U&=-e^{\frac{2\pi}{\beta}\left(r_*-t\right)}\,,\,\,V=+e^{\frac{2\pi}{\beta}\left(r_*+t\right)} \,\,\,(\text{right exterior region})  \\
U&=+e^{\frac{2\pi}{\beta}\left(r_*-t\right)}\,,\,\,V=+e^{\frac{2\pi}{\beta}\left(r_*+t\right)} \,\,\,(\text{future interior region}) \\
U&=-e^{\frac{2\pi}{\beta}\left(r_*-t\right)}\,,\,\,V=-e^{\frac{2\pi}{\beta}\left(r_*+t\right)} \,\,\,(\text{past interior region}) 
\end{split}
\end{equation}
where the tortoise coordinate is defined as
\be
 r_*(r)=\int^r \frac{dr'}{f(r')}\,,
\ee
and $\beta=4\pi/f'(r_0)$ is the black hole inverse temperature. 

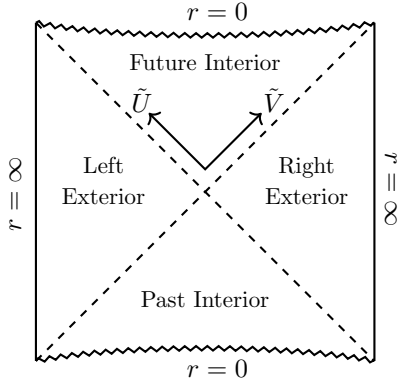
\begin{figure}[H]
\centering
%\captionsetup{justification=centering}

\begin{tikzpicture}[scale=1.5]
\draw [thick]  (0,0) -- (0,3);
\draw [thick]  (3,0) -- (3,3);
\draw [thick,dashed]  (0,0) -- (3,3);
\draw [thick,dashed]  (0,3) -- (3,0);
\draw [thick,decorate,decoration={zigzag,segment length=1.5mm, amplitude=0.3mm}] (0,3) .. controls (.75,2.85) 
and (2.25,2.85) .. (3,3);
\draw [thick,decorate,decoration={zigzag,segment length=1.5mm,amplitude=.3mm}]  (0,0) .. controls (.75,.15) and (2.25,.15) .. (3,0);

\draw[thick,<->] (1,2.2) -- (1.5,1.7) -- (2,2.2);

\node[scale=0.8, align=center] at (1.5,2.65) {Future Interior};
\node[scale=0.8,align=center] at (1.5,.55) {Past Interior};
\node[scale=0.8,align=center] at (0.6,1.6) {Left\\ Exterior};
\node[scale=0.8,align=center] at (2.4,1.6) {Right\\ Exterior};
\end{tikzpicture}
%\put(10,60){\Large $= |TFD \rangle$}
\vspace{0.1cm}
\put(-142,50){\rotatebox{90}{\small $r = \infty$}}
\put(-0,80){\rotatebox{-90}{\small $r = \infty$}}
\put(-75,-5){\small $r = 0$}
\put(-75,130){\small $r = 0$}
\put(-96,95){\small $\tilde{U}$}
\put(-46,95){\small $\tilde{V}$}

\caption{ \small Penrose diagram for two-sided black holes with asymptotically AdS geometry.}
\label{fig-Penrose}
\end{figure}

In terms of these coordinates, the metric reads
\be 
ds^2 = 2A(UV)dUdV+r^2(UV)dH_{d-1}^2\,,
\label{eq-metric-Kruskal}
\ee
where
\be
A(UV)=\frac{\beta^2}{8\pi^2}\frac{f(r(UV))}{UV}\,.
\ee
In these coordinates, the left and right asymptotic boundaries are located at $UV=-1$, and the past and future singularities at $UV=1$. One of the horizons is located at $U=0$, while the other one is located at $V=0$. The Penrose diagram\footnote{This diagram is obtained by an additional change of coordinates $U \rightarrow \tilde{U}=\tanh(U)$ and $V \rightarrow \tilde{V} =\tanh(V)$.} for this geometry is shown in figure \ref{fig-Penrose}.

\subsection{Rindler-AdS spacetime}
In embedding coordinates, the AdS$_{d+1}$ space is defined as the hyperboloid
\be
-T_1^2-T_2^2+X_1^2+...+X_d^2=-\ell^2\,,
\ee
with ambient metric
\be \label{metric123}
ds_{d+2}^2=-dT_1^2-dT_2^2+dX_1^2+...+dX_d^2\,.
\ee
The Rindler-AdS geometry  (also known as the ``Rindler wedge of AdS'' or as a ``topological black hole'') is defined as 
\begin{equation} \label{embed1}
\begin{split}
&T_1=\sqrt{r^2-\ell^2} \sinh \frac{t}{\ell} \,, \\
&T_2= r \cosh \chi \,, \\
&X_d=\sqrt{r^2-\ell^2} \cosh \frac{t}{\ell}  \,, \\
&X_1^2+...+X_{d-1}^2= r^2 \sinh^2 \chi \,,
\end{split}
\end{equation}
where $r \in [\ell,\infty)$, $\chi \in [0,\infty)$ and $t \in (-\infty,\infty)$. In terms of these coordinates, the metric becomes
\be
ds^2=-\left(\frac{r^2}{\ell^2}-1 \right)dt^2+\frac{dr^2}{\frac{r^2}{\ell^2}-1 }+r^2 dH_{d-1}^2\,.
\label{eq-metricR}
\ee
This corresponds to a special case of the metric (\ref{eq-metricS}), in which $r_0=\ell$. Note that in this case the Hawking inverse temperature becomes $\beta = 2 \pi \ell$.

For future purposes, it will also be useful to write the embedding coordinates in terms of Kruskal coordinates, namely
\begin{equation} \label{embed2}
\begin{split}
&T_1=\ell \frac{U+V}{1+UV} \,,  \\
&T_2=\ell \frac{1-UV}{1+UV} \cosh \chi  \,,  \\
&X_d=\ell \frac{V-U}{1+UV}  \,,  \\
&X_1^2+...+X_{d-1}^2= \ell^2 \left( \frac{1-UV}{1+UV} \right) \sinh^2 \chi \,,
\end{split}
\end{equation}
in terms of which the metric \eqref{metric123} becomes
\be
ds^2=-\frac{4 \ell^2 dU dV}{(1+UV)^2}+\left( \frac{1-UV}{1+UV} \right)^2 dH_{d-1}^2\,,
\ee
which corresponds to the metric \eqref{eq-metric-Kruskal} with $r_0=\ell$ or $\beta = 2\pi \ell$. 

\subsection{The dual CFT description}\label{sec-TFD}
The hyperbolic black hole geometry is dual to a CFT in hyperbolic space $\mathbb{R} \times H_{d-1}$. The maximally extended hyperbolic black hole geometry is dual to a thermofield double (TFD) state constructed by entangling two copies of such CFTs
\be
|\text{TFD} \rangle_{t=0}=\frac{1}{Z(\beta)^{1/2}} \sum_n e^{-\beta E_n/2} \,| E_n \rangle_\mt{L} \otimes | E_n \rangle_\mt{R}\,,\,\,\,\,\,\,\text{with}\,\,Z(\beta)=\text{Tr}\,e^{-\beta H}\,,\\
\ee
where each CFT has Hamiltonian $H$ and partition function $Z(\beta)$. Here, the subscript $L \,(R)$ denotes the energy eigenstates of the CFT living on the left (right) asymptotic boundary of geometry.

Interestingly, the pure $AdS_{d+1}$ geometry can be thought of as an entangled state of a pair of CFTs on hyperbolic space \cite{Czech:2012be}, with inverse temperature $\beta = 2 \pi \ell$. In this case, the corresponding geometry is simply the hyperbolic slicing of $AdS_{d+1}$, which is also known as the ``Rindler-AdS geometry''.  

\section{OTOCs from shock waves}\label{sec-shock}
\subsection{OTOCs in the eikonal approximation}

In this section, we use the elastic eikonal gravity approximation~\cite{BHchaos4} to compute OTOCs of the form
\be
F= \langle\text{TFD}|V_{{\bf x_1}}(t_1) W_{\bf x_2}(t_2) V_{\bf x_3}(t_3) W_{\bf x_4}(t_4) |\text{TFD}\rangle\,,
\ee
where $V$ and $W$ are single trace operators acting on the right side of the geometry. We regularize the OTOC by considering complex times
\begin{equation} \label{eq-times}
\begin{split}
t_1 &= -t/2+i \epsilon_1\,, \,\,\,\,\,\, t_3 = -t/2+i \epsilon_3\,, \\
t_2 &= t/2+i \epsilon_2\,, \,\,\,\,\,\,\,\,\,\,\, t_4 = t/2+i \epsilon_4\,.
\end{split}
\end{equation}

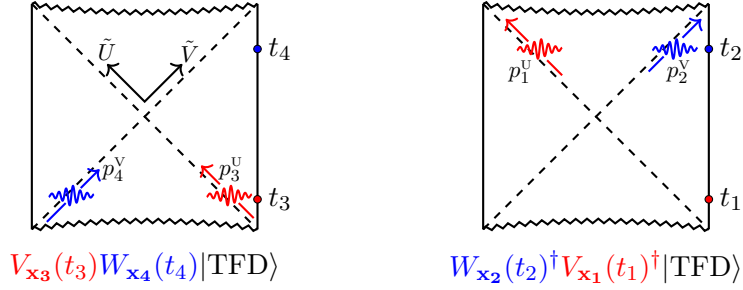
\begin{figure}[H]
\centering
%\captionsetup{justification=centering}
\begin{tikzpicture}[scale=1.]
\draw [thick]  (0,0) -- (0,3);
\draw [thick]  (3,0) -- (3,3);
\draw [thick,dashed]  (0,0) -- (3,3);
\draw [thick,dashed]  (0,3) -- (3,0);
\draw [thick,decorate,decoration={zigzag,segment length=1.5mm, amplitude=0.3mm}] (0,3) .. controls (.75,2.85) 
and (2.25,2.85) .. (3,3);
\draw [thick,decorate,decoration={zigzag,segment length=1.5mm,amplitude=.3mm}]  (0,0) .. controls (.75,.15) and (2.25,.15) .. (3,0);

%\draw [thick,blue] (3,2.8) -- (1.3,4.5);

%\node[scale=1,align=center] at (3.5,2.8){$W_{\phi_4}(t_4)$};

\node[scale=.8,align=center] at (1.0,2.40){$\tilde{U}$};
\node[scale=.8,align=center] at (2.1,2.37){$\tilde{V}$};

\draw[thick,<->] (1,2.2) -- (1.5,1.7) -- (2,2.2);

%  \draw[-,scale=.15,samples=300,thick,domain=-2:2] plot ({\x},{exp(-abs(\x))*sin(10*\x r)});

 %%%% wave function for psi_4 
  \draw [thick,blue] (0.2,0.1) -- (.45,.35);
  \draw [->,thick,blue] (0.65,0.55) -- (.9,.8);
  \draw[-,blue,scale=.15,samples=300,thick,domain=1.5:5.5] plot ({\x},{3+exp(-abs(\x-3.5))*sin(10*(\x-3.5) r)});
  \node[scale=.8,align=center] at (1.1,.8){$p_4^\mt{V}$};

 %%%% wave function for psi_3 
 \draw[-,red,scale=.15,samples=300,thick,domain=15.5:19.5] plot ({\x},{3+exp(-abs(\x-17.5))*sin(10*(\x-17.5) r)});
\draw [thick, red] (2.95,0.15) -- (2.75,0.35);
\draw [->,thick, red] (2.5,0.6) -- (2.25,0.85);
 \node[scale=.8,align=center] at (2.65,.8){$p_3^\mt{U}$};

\node[scale=1,align=center] at (1.5,-0.5){${\color{red}{V_{\bf x_3}(t_3)}} {\color{blue}{W_{\bf x_4}(t_4)}}| \text{TFD}\rangle$};

\draw [fill=red] (3,.4) circle [radius=0.05];
 \node[scale=1,align=center] at (3.3,.4){$t_3$};

\draw [fill=blue] (3,2.4) circle [radius=0.05];
 \node[scale=1,align=center] at (3.3,2.4){$t_4$};

\draw [thick]  (6,0) -- (6,3);
\draw [thick]  (9,0) -- (9,3);
\draw [thick,dashed]  (6,0) -- (9,3);
\draw [thick,dashed]  (6,3) -- (9,0);
\draw [thick,decorate,decoration={zigzag,segment length=1.5mm, amplitude=0.3mm}] (6,3) .. controls (6.75,2.85) 
and (8.25,2.85) .. (9,3);
\draw [thick,decorate,decoration={zigzag,segment length=1.5mm,amplitude=.3mm}]  (6,0) .. controls (6.75,.15) and (8.25,.15) .. (9,0);

\draw [fill=red] (9,.4) circle [radius=0.05];
 \node[scale=1,align=center] at (9.3,.4){$t_1$};

\draw [fill=blue] (9,2.4) circle [radius=0.05];
 \node[scale=1,align=center] at (9.3,2.4){$t_2$};
 
 \node[scale=1,align=center] at (7.5,-0.5){${\color{blue}{W_{\bf x_2}(t_2)^{\dagger}}} {\color{red}{V_{\bf x_1}(t_1)^{\dagger}}}| \text{TFD}\rangle$};

 %%%% wave function for psi_1 
 \draw[-,red,scale=.15,samples=300,thick,domain=43:47] plot ({\x},{16+exp(-abs(\x-45))*sin(10*(\x-45) r)});
\draw [->,thick, red] (6.6,2.5) -- (6.3,2.8);
\draw [thick, red] (6.85,2.25) -- (7.05,2.05);

\node [scale=.8,align=center] at (6.5,2.1){$p_1^\mt{U}$};

 %%%% wave function for psi_4 
 \draw [thick,blue] (8.4,2.3) -- (8.2,2.1);
  \draw [<-,thick,blue] (8.9,2.8) -- (8.65,2.55);
  \draw[-,blue,scale=.15,samples=300,thick,domain=55:59] plot ({\x},{16+exp(-abs(\x-57))*sin(10*(\x-57) r)});
 \node[scale=.8,align=center] at (8.6,2.1){$p_2^\mt{V}$};

\end{tikzpicture}

\caption{ \small {\it Left}: representation of the `in' state ${\color{red}{V_{\bf x_3}(t_3)}} {\color{blue}{W_{\bf x_4}(t_4)}}| \text{TFD}\rangle$ on a bulk slice that touches the right boundary at time $t_3$. {\it Right}: representation of the `out' state ${\color{blue}{W_{\bf x_2}(t_2)^{\dagger}}} {\color{red}{W_{\bf x_1}(t_1)^{\dagger}}}| \text{TFD}\rangle$ on a bulk slice that touches the right boundary at time $t_2$.}
\label{fig-In-Out-states}
\end{figure}

Following~\cite{BHchaos4}, we write the OTOC as a scattering amplitude
\be
F = \langle \text{out} |  \text{in} \rangle\,,
\ee
where $| \text{in} \rangle = V_{\bf x_3}(t_3) W_{\bf x_4}(t_4) |\text{TFD}\rangle$ and $| \text{out} \rangle = W_{\bf x_2}(t_2)^{\dagger} V_{\bf x_1}(t_1)^{\dagger} |\text{TFD}\rangle$ are `in' and `out' states. In the bulk, these states can be described in terms of two particle states, which can be represented on any bulk slice. See figure \ref{fig-In-Out-states}. We call $V-$particle ($W-$particle) the field excitation dual to the operator $V$ ($W$). We will be interested in the configuration where $t$ is large. In this case the $V-$particle ($W-$particle) will be highly boosted with respect to the $t=0$ slice of the geometry, having a large momentum in the $V-$direction ($U-$direction). The `in' state represents the $V$ and $W$ particles heading to collide, while the `out' state represents the outcome of that collision.

For convenience, we decompose the state of the $V-$particle in the basis $|p^\mt{U},{\bf x} \rangle$ of well-defined momentum and  position, and represent it in the $U=0$ slice of the geometry. In the same way, we decompose the state of the $W-$particle in the basis $|p^\mt{V},{\bf x'} \rangle$ and represent it in the $V=0$ slice of the geometry.
By representing $V$ and $W$ via the `extrapolate' dictionary, we write the `in' state as
\be
V_{\bf x_3}(t_3) W_{\bf x_4}(t_4) |\text{TFD}\rangle = \int d {\bf x_3'} \, d {\bf x_4'} \int dp_3^\mt{U} dp_3^\mt{V} \, \psi_3(p_3^\mt{U} ,{\bf x_3'} ) \psi_4(p_4^\mt{V} ,{\bf x_4'} ) |p_3^\mt{U} ,{\bf x_3'} \rangle \otimes | p_4^\mt{V} ,{\bf x_4'} \rangle\,,
\ee
while the `out' state is written as
\be
V_{\bf x_1}(t_1)^{\dagger} W_{\bf x_2}(t_2)^{\dagger} |\text{TFD}\rangle = \int d {\bf x_1'} \, d{\bf x_2'} \int dp_1^\mt{U} dp_2^\mt{V} \, \psi_1(p_1^\mt{U} ,{\bf x_1'} ) \psi_2(p_2^\mt{V} ,{\bf x_2'} ) |p_1^\mt{U} ,{\bf x_1'} \rangle \otimes | p_2^\mt{V} ,{\bf x_2'} \rangle\,.
\ee

The wave functions $\psi_i$ featuring in the above formulas are Fourier transforms of bulk-to-boundary propagators along either the $U=0$ or $V=0$ horizons
\begin{equation}
\begin{split}
\psi_1(p^\mt{U},{\bf x}) &= \int  dV e^{i A_0 p^\mt{U} V} \langle \Phi_V(U,V, {\bf x} ) V_{\bf x_1}(t_1)^{\dagger}\rangle|_{U=0} \,, \\
\psi_2(p^\mt{V},{\bf x}) &= \int  dU e^{i A_0 p^\mt{V} U} \langle \Phi_W(U,V, {\bf x} ) W_{\bf x_2}(t_2)^{\dagger}\rangle|_{V=0} \,,\\
\psi_3(p^\mt{U},{\bf x}) &= \int  dV e^{i A_0 p^\mt{U} V} \langle \Phi_V(U,V, {\bf x}) V_{\bf x_3}(t_3) \rangle|_{U=0} \,,\\
\psi_4(p^\mt{V},{\bf x}) &= \int  dU e^{i A_0 p^\mt{V} U} \langle \Phi_W(U,V, {\bf x} ) W_{\bf x_4}(t_4) \rangle|_{V=0} \,,
\end{split}
\end{equation}
where the bulk fields $\Phi_V$ and $\Phi_W$ are dual to the operators $V$ and $W$. 

The measure factors are given by
\be \label{defx1}
d {\bf x} =\sinh^{d-2}\chi \, d \chi \,d \Omega_{d-2},
\ee
with $d \Omega_{d-2}=\sin^{d-3}\theta_{d-3} \,\, \cdots \,\, \sin \theta_1 d \phi d\theta_1 \cdots d\theta_{d-2}$.  We normalize the basis vectors as
\be
\langle p^\mt{U},{\bf x} | q^\mt{U},{\bf x'} \rangle = \frac{A_0^2 \, p^\mt{U}}{\pi r_{0}^{d-1}} \delta(p^\mt{U}-q^\mt{U})\, \delta({\bf x,x'})\,,
\ee
where we defined $A_0 := A(0)$ and $\delta({\bf x,x'}) := \frac{\delta(\chi-\chi')}{\sinh^{d-2} \chi} \frac{ \delta(\theta_1-\theta_1') \,\,\cdots \,\, \delta(\theta_{d-2}-\theta_{d-2}')}{\sin \theta_1'\,\, \cdots \,\,\sin^{d-3}\theta'_{d-3}} \delta(\phi-\phi')$.

The collision takes place close to the bifurcation surface (at $U=V=0$), where both particles have very large momentum.  In this configuration, since the collision impact parameter (denoted by $b$) is fixed and $G_{N}$ is small, the gravitational interaction dominates over all other interactions, and the amplitude is dominated by ladder and crossed ladder diagrams involving graviton exchanges~\cite{Kabat:1992tb}. This leads to the very simple result
\be
\left( |p_1^\mt{U} ,{\bf x_1} \rangle  \otimes | p_2^\mt{V} ,{\bf x_2} \rangle \right)_{\text{out}} \approx e^{i \delta(s,b)} \left( |p_1^\mt{U} ,{\bf x_1} \rangle \otimes | p_2^\mt{V} ,{\bf x_2} \rangle \right)_\text{in}+| \text{inelastic}\rangle\,,
\ee
where the phase shift $\delta(s,b)$ depends on $s=(p_1+p_2)^2=2A_0 \,p^\mt{U}p^\mt{V}$ and $b$ is the impact parameter.
The state $| \text{inelastic}\rangle$ accounts for an inelastic contribution that is orthogonal to all two-particle `in' states.

Using the above formulas, the OTOC can be written as
\be
F= \frac{A_0^4}{\pi^2} \int \int d {\bf x} d {\bf x'} \int \int dp_1^\mt{U} dp_2^\mt{V} e^{i\delta(s,b)} \Big[ p_1^\mt{U} \psi_1^*(p_1^\mt{U},{\bf x}) \psi_3(p_1^\mt{U},{\bf x}) \Big] \Big[p_2^\mt{V} \psi_2^*(p_2^\mt{V},{\bf x'}) \psi_4(p_2^\mt{V},{\bf x'}) \Big]\,.
\label{eq-OTOCformula}
\ee
Thus, once we know the phase shift $\delta(a,b)$ and the wave functions $\psi_i$ we can compute the OTOC. In the next subsection, we explain how to compute the phase shift for general hyperbolic black holes, with metric of the form (\ref{eq-metricS}). The computation of the wave functions $\psi_i$ requires the knowledge of bulk-to-boundary propagators, which are unknown for general hyperbolic black holes. However, for the special case of a Rindler-AdS geometry, which can be obtained from (\ref{eq-metricS}) by setting $r_0=\ell$, the bulk-to-boundary propagators are known, and the wave functions can be computed. In this case, (\ref{eq-OTOCformula}) can be evaluated, and one obtains an analytic result for the OTOCs. This calculation is done in subsection  \ref{btb123}. The case of general hyperbolic black holes, in which (\ref{eq-OTOCformula}) cannot be evaluated precisely, is discussed in section \ref{sec-otocsGeneral}.

\subsubsection{The phase shift}\label{sec-phase}
In the elastic eikonal gravity approximation, the phase shift is given by 
\be
\delta(s,b) = S_\text{classical}\,,
\ee
where $S_\text{classical}$ is the sum of the on-shell actions for the $V$ and $W$ particles. To compute this action, we need to know the stress-energy tensor of these particles, and the corresponding back-reaction on the geometry. 

For very large $t$, the $V-$particle follows an almost null trajectory, very close to the $V=0$ horizon. In this configuration, the stress-energy
of this particle reads
\be
V-\text{particle}: T_{VV}({\bf x,x'})=\frac{A_0}{r_{0}^{d-1}}p_1^\mt{U} \delta(V) \delta({\bf x,x'})\,,
\ee
where ${\bf x'}$ denotes the position of the $V-$particle in $H_{d-1}$. The corresponding back-reaction on the geometry can be simply obtained with the replacement
\be
ds^2 \rightarrow ds^2 + h_{VV} dV^2\,,\,\,\,\, h_{VV}= \frac{16 \pi G_N  A_0}{r_0^{d-1}} p_1^\mt{U} \delta(V) h(d({\bf x,x'}))\,,
\ee
where $ds^2$ denotes the unperturbed geometry (\ref{eq-metric-Kruskal}), and the shock wave transverse profile $h(d({\bf x,x'}))$ is a solution of the equation
\be
\left[ \square_{H_{d-1}}-\frac{2\pi}{\beta} r_0 (d-1) \right] h(d({\bf x,x'}))=-\frac{8 \pi G_{N}}{r_0^{d-3}} p^\mt{V}\delta({\bf x,x'})\,.
\label{eq-shockwaveprofile}
\ee
Here, the function $h$ is a function of $d({\bf x,x'})$, which is the geodesic distance between $\bf x$ and $\bf x'$ in $H_{d-1}$. Its explicit form is given in \eqref{eq-hypdist}.

For large values of $d({\bf x, {x'}})$, the shock wave transverse profile behaves as\footnote{See Appendix \ref{AppA} for more details.}
\be \label{h1234}
h(\chi) = c_1\, e^{-\mu  d({\bf x, {x'}})}\,,\,\,\,\,\,\, \mu \equiv \frac{1}{2}\left(d-2+\sqrt{(d-2)^2+\frac{8\pi r_0}{\beta}(d-1)} \right)\,,
\ee
where $c_1$ is a constant.

The $W-$particle, by its turn, follows an almost null trajectory very close to the $U=0$ horizon, with stress-energy tensor given by
\be
W-\text{particle}: T_{UU}({\bf x,x''})=\frac{A_0}{r_{0}^{d-1}}p_2^\mt{V} \delta(U) \delta({\bf x,x''}) \,.
\ee
The corresponding back-reaction on the geometry is obtained with the replacement
\be
ds^2 \rightarrow ds^2 + h_{UU} dU^2\,,\,\,\,\, h_{UU}= \frac{16 \pi G_N  A_0}{r_0^{d-1}} p_2^\mt{V} \delta(U) h(d({\bf x,x''}))\,.
\ee

The on-shell action can be written as~\cite{BHchaos4,Kabat:1992tb}
\be
S_\text{classical} =\frac{1}{4}\int d^{d+1}x \sqrt{-g} \left( h_{UU}T^{UU}+ h_{VV}T^{VV} \right)\,.
\ee
The above formula is actually symmetric in the exchange of the two particles: while $h_{UU}$ refers to the $W-$particle, the stress-energy tensor $T^{UU}=g^{UV}g^{UV}T_{VV}$ refers to the $V-$particle, with a similar story for $h_{VV}$ and $T^{VV}$.  Substituting the expressions for the stress-energy tensors and the corresponding back-reactions, we find
\be
\delta(s,b)= \frac{4\pi G_N }{r_0^{d-1}}s\,h(b/\ell)   =  \frac{8\pi G_N }{r_0^{d-1}} A_0 \,p_1^\mt{U} p_2^\mt{V}\, c_1 \,e^{-\frac{\mu}{\ell} b}\,,
\label{eq-phase}
\ee
where  $s=2A_0 \,p_1^\mt{U} p_2^\mt{V}$ and $b=\ell \,d({\bf x',x''})$ is an impact parameter of length dimension while the geodesic distance $d({\bf x',x''})$ is dimensionless.  We emphasize that (\ref{eq-phase}) is valid for a generic hyperbolic black hole, as long as the metric has the form (\ref{eq-metricS}).

\subsection{OTOCs in the Rindler-AdS$_{d+1}$ geometry}\label{btb123}

In this section, we compute OTOCs for a Rindler-AdS$_{d+1}$ geometry. We start by computing the geodesic distance between points in this geometry. From this we can easily obtain the bulk-to-boundary propagators and then the wave functions $\psi_i$, which are essential ingredients for evaluating (\ref{eq-OTOCformula}).

%We compute bulk-to-boundary propagators in AdS-Rindler space by considering the geodesic distance between points in this space. 

First, the geodesic distance $d(p,p')$ between two points $p=(T_1,T_2,X_1,...,X_d)$ and $p'=(T_1',T_2',X_1',...,X_d')$ is~\cite{BHchaos1}
\be \label{geod1}
\cosh \left(  \frac{d(p,p')}{\ell} \right) =\frac{1}{\ell^2}\left( T_1 T_1' + T_2 T_2' - X_1 X_1' -X_2 X_2'-...-X_d X_d'\right) \,.
\ee
It is convenient to write the boundary point in terms of AdS-Rindler coordinates $p'=(t,r,{\bf x}')$ (in the limit $r \rightarrow \infty$) and the bulk point in terms of Kruskal coordinates $p=(U,V,{\bf x})$. Here ${\bf x}=(\chi,\Omega_{d-2})$ and ${\bf x'}=(\chi',\Omega'_{d-2})$ denote points in hyperbolic space $H_{d-1}$, with $\Omega_{d-2}$ and $\Omega'_{d-2}$ being points in the sphere $S^{d-2}$. 

Eq.~\eqref{geod1}  can then be written as\footnote{Here, to simplify our formulas and avoid clutter, we set $\ell=1$. This fixes the inverse Hawking temperature as $\beta=2\pi$.}
\be
\cosh d(p,p') =\frac{1}{1+UV} \left[\sqrt{r^2-1}\left( U e^t -V e^{-t} \right)+r (1-UV) \,\cosh d({\bf x,x'}) \right] \,,
\ee
where \eqref{embed1} and \eqref{embed2} were used
and  $d({\bf x,x'})$ is the geodesic distance between the points ${\bf x}$ and ${\bf x'}$ in $H_{d-1}$. This distance can be written as~\cite{Cohl:2012}
\be
d({\bf x,x'})=\cosh^{-1} \left( \cosh \chi \cosh \chi' - \sinh \chi \sinh \chi' \cos \gamma \right) \,,
\label{eq-hypdist}
\ee
with
\be
\cos \gamma := \cos(\phi-\phi') \prod_{i=1}^{d-3} \sin \theta_i \sin \theta_i'+\sum_{i=1}^{d-3} \cos \theta_i \cos \theta_i' \prod_{j=1}^{i-1} \sin \theta_j \sin \theta_j'  \,.
\ee 
Here $\gamma$ may be understood as  the geodesic distance between two points $\Omega_{d-2}$ and $\Omega'_{d-2}$  in the sphere $S^{d-2}$.
Here, $\phi \in [0,2\pi)$ and $\theta_i \in [0,\pi]$. For example, in $S^2$, $\cos \gamma =\cos(\phi-\phi') \sin \theta \sin \theta'+\cos \theta \cos \theta'$, where $\theta \in [0,\pi]$ is the polar angle, while $\phi \in [0,2\pi)$ is the azimuthal angle.

Having computed the geodesic distances, the bulk-to-bulk propagator associated to a bulk field $\Phi$, dual to an operator $\mathcal{O}_{\Delta}$ of scaling dimension $\Delta$, can be obtained as~\cite{Ammon:2015wua}
\be
G_{\Delta}(p ; p') = \frac{\Gamma(\Delta)}{\pi^{d/2}\Gamma(\Delta-\frac{d}{2})} \left( \cosh d(p,p') \right)^{-\Delta} \,.
\ee
The bulk-to-boundary propagator can then be computed as~\cite{Ammon:2015wua}
\be
\langle \Phi(U,V,{\bf x}) \mathcal{O}_{\Delta}( t,{\bf x}') \rangle = \left(2\Delta-1 \right) \lim_{r\rightarrow \infty} r^{\Delta }[G_{\Delta}(U,V,{\bf x} ; t,r,{\bf x}')]\,.
\ee
Using the above formulas, we find
\be
\langle \Phi(U,V,{\bf x}) \mathcal{O}_{\Delta}( t,{\bf x}') \rangle = C_{\Delta} \,  \Big[U e^{t}-V e^{-t}+(1-UV) \cosh d({\bf x,x_1}) \Big]\,,
\label{eq-bulkboundaryprop} 
\ee
where $C_{\Delta}=\frac{\Gamma(\Delta)}{\pi^{d/2}\Gamma(\Delta-\frac{d}{2})}\,$. 

We are now ready to evaluate the integral (\ref{eq-OTOCformula}) for an Rindler-AdS$_{d+1}$ geometry. Since we set $\ell=1$, we have $A_0=2$, $r_0=1$ and $\beta=2\pi$. Using (\ref{eq-bulkboundaryprop}), the bulk-to-boundary propagators can be written as 
\bea
& &\langle \Phi_V(U,V,{\bf x}) V_{\bf x_1}(t_1)\rangle = c_\mt{V}\, \Big[U e^{t}-V e^{-t}+(1-UV) \cosh d({\bf x,x_1}) \Big]^{-\Delta_V}\,,\\
& &\langle  \Phi_W(U,V,{\bf x'}) W_{\bf x_2}(t_2)\rangle = c_\mt{W}\, \Big[ U e^{t}-V e^{-t}+ (1-UV) \cosh d({\bf x',x_2}) \Big]^{-\Delta_W}\,,
\eea
from which we obtain the following wave functions
\begin{equation}
\begin{split}
\psi_1(p^\mt{U},{\bf x})&=-\theta(p^\mt{U}) \frac{2\pi i c_V}{\Gamma(\Delta_V)}    e^{ t_1^\mt{*}} \left(-2i p^\mt{U} e^{ t_1^\mt{*}} \right)^{\Delta_V-1} e^{2i  p^\mt{U} e^{ t_1^\mt{*}} \cosh d({\bf x,x_1})  }\,,  \\
\psi_3(p^\mt{U},{\bf x})&=-\theta(p^\mt{U}) \frac{2\pi i c_V}{\Gamma(\Delta_V)}    e^{ t_3} \left(-2i  p^\mt{U} e^{ t_3} \right)^{\Delta_V-1} e^{2i  p^\mt{U} e^{ t_3} \cosh d({\bf x,x_1}) }\,,   \\
\psi_2(p^\mt{V},{\bf x'})&=\theta(p^\mt{V}) \frac{2\pi i c_W}{\Gamma(\Delta_W)}    e^{ -t_2^\mt{*}} \left(2i p^\mt{V} e^{ -t_2^\mt{*}} \right)^{\Delta_W-1} e^{-2i  p^\mt{V} e^{-t_2^\mt{*}} \cosh d({\bf x',x_2})  } \,,  \\
\psi_2(p^\mt{V},{\bf x'})&=\theta(p^\mt{V}) \frac{2\pi i c_W}{\Gamma(\Delta_W)}    e^{ -t_4} \left(2i p^\mt{V} e^{ -t_4} \right)^{\Delta_W-1} e^{-2i  p^\mt{V} e^{-t_4} \cosh d({\bf x',x_2})  } \,.
\end{split}
\end{equation}

Using the above formulas, the OTOC becomes
\be
F=  K \int d{\bf x} d{\bf x'} dp^\mt{U} dp^\mt{V} e^{\delta(s,b({\bf x,x'}))} (p^\mt{U})^{2\Delta_V-1} (p^\mt{V})^{2\Delta_W-1} 
 \frac{e^{ \Delta_V  (t_1+t_3)}}{e^{ \Delta_W  (t_2+t_4)}}  \frac{e^{2i p^\mt{U} \cosh d({\bf x,x_1}) \left( e^{ t_3 } - e^{ t_1} \right) }}{ e^{2i p^\mt{V} \cosh d({\bf x',x_2}) \left( e^{t_4 } - e^{t_2} \right) }}\,,
\ee
where $K:=\left( \frac{4\pi c_V c_W 2^{\Delta_V+\Delta_W}}{\Gamma(\Delta_V) \Gamma(\Delta_W)} \right)^2$. By introducing the new variables 
\begin{equation}
\begin{split}
p&= -2i  p^\mt{U} \left( e^{ t_3}-e^{ t_1}\right)\,, \\
q&=  2i  p^\mt{V} \left( e^{ t_4}-e^{ t_2}\right)\,,
\end{split}
\end{equation}
and specifying the times as in (\ref{eq-times}), the integral becomes
\be
F=C \int d{\bf x} d{\bf x'} dp dq  p^{2\Delta_V-1} q^{2\Delta_W-1} e^{-p \cosh d({\bf x,x_1})} e^{-q \cosh d({\bf x',x_2})} e^{ i G_N p\,q \,e^t h(d({\bf x,x'}))/\epsilon_{13}\epsilon_{24}^\mt{*}}\,,
\ee
where $\epsilon_{ij}:=i(e^{ i \epsilon_i}-e^{i \epsilon_j})$ and $C$ is a constant given by
\be
C=\frac{2 \pi^2 c_V^2 c_W^2}{\Gamma(\Delta_V)^2 \Gamma(\Delta_W)^2} \left[ \frac{1}{2 \sin\left( \frac{\epsilon_3-\epsilon_1}{2}\right)}\right]^{2\Delta_V} \left[ \frac{1}{2 \sin\left( \frac{\epsilon_4-\epsilon_2}{2}\right)}\right]^{2\Delta_W}\,.
\ee

If we set $\delta(s,b)=0$, the above integral gives $\langle V V\rangle \langle W W \rangle$. For $\delta(s,b)\neq0$, the integral can be evaluated in the limit $\Delta_W \gg \Delta_V \gg 1$ and the result reads\footnote{Here, we first write the integrals in ${\bf x'}$ and $q$ in the form $\int d{\bf x'} dq\, e^{-F({\bf x'},q)}$ and check that the result is dominated by the region of integration where $d({\bf x',x_2}) \approx 0$ and $q \approx 2 \Delta_W$. After this, the integral in $p$ can be done analytically, and the integral in ${\bf x}$ can be done by a saddle point approximation.}
\be
\frac{\langle V_{\bf x_1}(t_1) W_{\bf x_2}(t_2) V_{\bf x_1}(t_3) W_{\bf x_2}(t_4) \rangle }{\langle V_{\bf x_1}(i \epsilon_1) V_{\bf x_1}(i \epsilon_3)\rangle \langle W_{\bf x_1}(i \epsilon_2) W_{\bf x_4}(i \epsilon_4) \rangle}=\frac{1}{\left[ 1- \frac{16 \pi\, i\, G_N \Delta_{W}}{ \epsilon_{13} \epsilon^\mt{*}_{24}}e^{ t}h(d({\bf x_1,x_2}))\right]^{\Delta_V}} \,. 
\ee
By writing $t_*=\log \frac{1}{16 \pi G_N}$  and using that $h(d({\bf x_1,x_2}))= c_1\, e^{-(d-1)d({\bf x_1,x_2})}$ (see Appendix \ref{AppA}) we can see that, for $t \lesssim t_*$, the OTOC behaves as
\be \label{otoc123}
\text{OTOC}(t,b)=1+\frac{ i\,c_1 \Delta_{W} \Delta_V}{ \epsilon_{13} \epsilon^\mt{*}_{24}}e^{ t-t_*-(d-1)b}\,,\,\,\,\,\,b := d({\bf x_1,x_2})\,,
\ee
from which we can extract the Lyapunov exponent $\lambda_L=\frac{2\pi}{\beta}=1$\footnote{Recall that $\beta = 2\pi$ in the Rindler-AdS geometry.}, and the butterfly velocity $v_B=\frac{1}{d-1}$.    This result matches the CFT result obtained by Perlmutter in~\cite{Perlmutter:2016pkf}.

\subsection{OTOCs in general hyperbolic black holes }\label{sec-otocsGeneral}
In this section, we consider general hyperbolic black holes, with a metric of the form (\ref{eq-metricS}). In these cases, the bulk-to-boundary propagators are unknown, so we cannot evaluate the integral (\ref{eq-OTOCformula}). We can, however, proceed as in~\cite{BHchaos4} and focus on the phase shift $\delta(s,b)$, which essentially controls the magnitude of the OTOC.

In section \ref{sec-phase}, we show that the phase shift is given by \eqref{eq-phase}:
\be
\delta(s,b)= \frac{8\pi G_N }{r_0^{d-1}} A_0 \,p_1^\mt{U} p_2^\mt{V}\, c_1 \,e^{-\frac{\mu}{\ell} b}\,,
\ee
where
\be
\mu = \frac{1}{2}\left(d-2+\sqrt{(d-2)^2+\frac{8\pi r_0}{\beta}(d-1)} \right)\,.
\ee
Let us assume that $W_{\bf x_2}(t/2)$ and $V_{\bf x_1}(-t/2)$ are thermal scale operators that raise the energy of the thermal state by an amount of order of the temperature $T$. That means that, when time equals $-t/2$, the $V$ particle is close to the boundary and has momentum $p_1^\mt{U} \approx T$. The $W$ particle, by its turn, is close to the boundary when time equals $t/2$, having momentum $p_2^\mt{V} \approx T$. The collision, however, takes place near the bifurcation surface, at the $t=0$ slice of the geometry. In this time slice, we have $p_1^\mt{U} \approx p_2^\mt{V} \approx T e^{\frac{\pi}{\beta}t}$, because the momentum of the $V-$particle increases exponentially as it falls into the black hole, while the momentum of the $W-$particle decreases exponentially as it escapes from the near-horizon region\footnote{See, for instance,~\cite{Susskind:2018tei}.}. 

This implies that, close to the bifurcation surface (at $U=V=0$), we have 
\be
p_1^\mt{U} p_2^\mt{V} \approx T^2 e^{\frac{2\pi}{\beta}t}\,.
\ee
With the above result the phase shift becomes
\be
\delta(s,b) \approx \frac{8\pi G_N }{r_0^{d-1}} A_0 T^2 e^{\frac{2\pi}{\beta}t-\frac{\mu}{\ell} b}\,, \,\,\,\, b := \ell \,d({\bf x_1,x_2}),
\ee
from which we can extract the maximal Lyapunov exponent $\lambda_L=\frac{2\pi}{\beta}$ and the butterfly velocity
\be
v_B(\scriptr_\mt{0}) \equiv \frac{2\pi \ell}{\beta \mu} = \frac{\sqrt{d \,\big[ 2\scriptr_\mt{0}^2 (d-1)-(d-2)\big] }-(d-2)}{2(d-1) \scriptr_\mt{0}}\,,\,\,\,\,\,      \scriptr_\mt{0}:=\frac{r_0}{\ell}\,.
\label{eq-vb}
\ee
In section \ref{sec-poleskip}, we obtain the same result for $\lambda_L$ and $v_B$ using a pole-skipping analysis.

\begin{figure}
\centering
 \includegraphics[width=8.5cm]{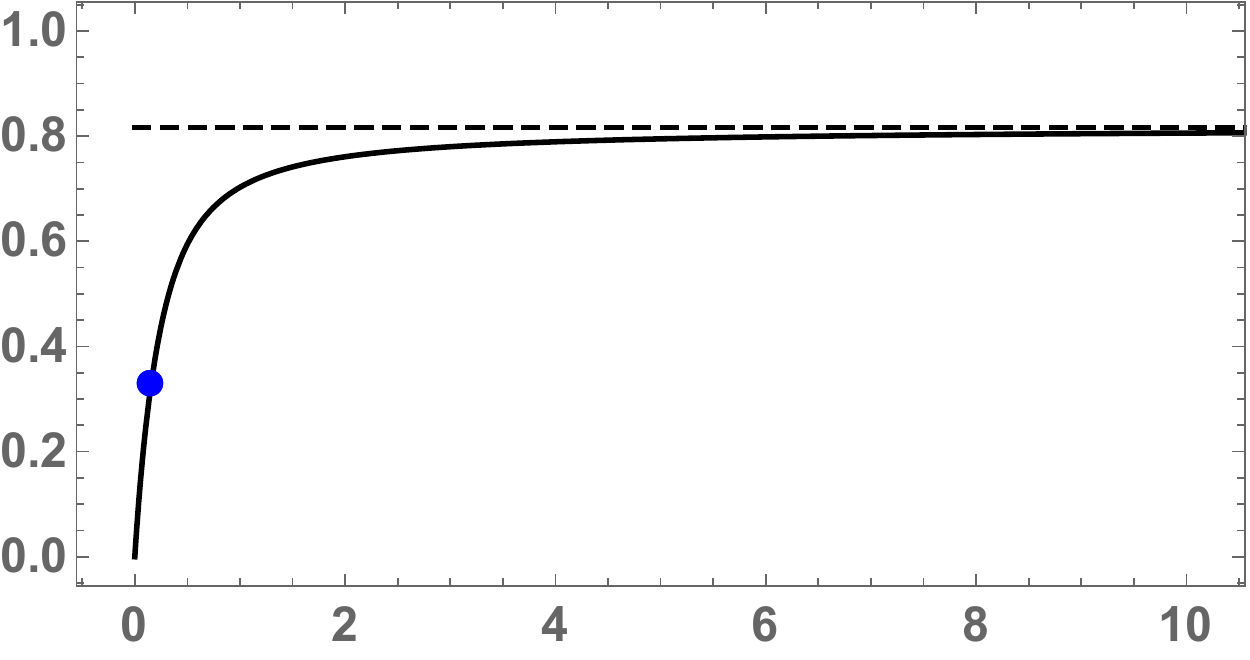} 
 \put(-260,70){\large $v_B$}
  \put(-110,-10){\large $T$}
  \put(5,100){\large $\sqrt{\frac{d}{2(d-1)}}$}
\caption{Temperature dependence of the butterfly velocity for hyperbolic black holes. The Rindler-AdS$_{d+1}$ result $(T,v_B)=(\frac{1}{2\pi \ell},\frac{1}{d-1})$ is indicated by the blue dot. Here we set $d=4$ and $\ell=1$. } 
\label{fig-vb1}
\end{figure}

The result (\ref{eq-vb}) has some interesting limits
\begin{itemize}
\item $v_B(\scriptr_\mt{0}=1)=\frac{1}{d-1}$, which is (as expected) the result for Rindler-AdS$_{d+1}$. By naively applying the  formula   derived for planar black holes, $v_B=\sqrt{\frac{2\pi \ell^2}{\beta(d-1)r_0}}$ (see~\cite{BHchaos4}, page 18), one gets the wrong result $v_B=\frac{1}{\sqrt{d-1}}$, which differs from the correct one by a square root;
\item $v_B(\scriptr_\mt{0} \gg 1)=\sqrt{\frac{d}{2(d-1)}}$, which is the result for a very large black hole $(r_0 \gg \ell)$. In this case, the butterfly velocity takes the planar value, i.e., the value for a $d-$dimensional CFT in flat space;
\item $v_B(\scriptr_\mt{0}=\scriptr_c)=0$, where $\scriptr_c=\sqrt{\frac{d-2}{d}}$. This happens because $v_B = \frac{2\pi \ell T}{\mu}$, and the black hole's temperature is zero for $\scriptr_\mt{0}=\scriptr_c$;
\item For $\scriptr_c \leq \scriptr_\mt{0} \leq 1$, the temperature is positive, but black hole's mass $M$ is negative ($M =\frac{(d-1)\text{vol}(H_{d-1})}{16 \pi G_N} (\scriptr_\mt{0}^2-1)$). 
In this case, the emblackening factor $f(r)$ has two distinct zeros, giving rise to an inner and an outer horizon. The Penrose diagram of the negative-mass black holes is similar to that of rotating black holes \cite{Mann:1997iz, Brill:1997mf, Banados:1992gq}\footnote{See \cite{Mann:1997iz, Brill:1997mf} for the negative-mass black holes case and figure 4 of \cite{Banados:1992gq} for the rotating black hole case.}.
Furthermore the formalism we used to compute OTOCs still applies, because the bulk description of two-particle states in terms of shock waves is essentially the same as the rotating black hole case.
For more details about why our derivation still applies, we refer to section 3 of \cite{Jahnke:2019gxr}, where OTOCs  were derived for rotating BTZ black holes. Interestingly, the butterfly velocity does not show any pathological behavior in this range.
\end{itemize}
The temperature behavior of the butterfly velocity can be obtained by writing the Hawking temperature as
\be
T(\scriptr_\mt{0},\ell)=\frac{2+d(\scriptr_\mt{0}^2-1)}{4\pi \ell \scriptr_\mt{0}} \,,
\label{eq-T}
\ee
and then making a parametric plot of $v_B(\scriptr_\mt{0})$ versus $T(\scriptr_\mt{0},\ell)$. This is shown in figure \ref{fig-vb1}, where we can see that $v_B$ starts at zero at $T=0$, increases as we increase $T$, and approaches the planar value $\sqrt{\frac{d}{2(d-1)}}$ for $T \gg1/\ell$.

\section{The pole-skipping analysis}\label{sec-poleskip}
In addition to OTOCs, the chaotic nature of many-body thermal systems is also encoded in energy density two-point functions. These functions exhibit a curious behavior, referred to as {\it pole-skipping}, from which one can extract both the Lyapunov exponent and the butterfly velocity of the system~\cite{Grozdanov:2017ajz,Blake:2017ris}. In this section, we use the pole-skipping analysis proposed in~\cite{Blake:2018leo} to extract the chaotic properties of 4-dimensional hyperbolic black holes.

\subsection{Pole-skipping: a brief review}

In momentum space, a generic retarded two-point function can be written as
\be \label{qnm1}
G^{R}(\omega,k)=\frac{b(\omega,k)}{a(\omega,k)}\,.
\ee
The poles of $G^R$ are generically described by a dispersion relation of the form $\omega=F(k)$, which corresponds to the zeros of $a(\omega,k)$. The pole-skipping phenomenon refers to the existence of special points, $(\omega_*,k_*)$, satisfying the following conditions
\be
\begin{split}
a(\omega_*,k_*)& =0\,, \\
b(\omega_*,k_*)&=0\,.
\end{split}
\ee
The first equation implies that the curve $\omega=F(k)$ passes through the special point $(\omega_*,k_*)$, while the second equation implies that $(\omega_*,k_*)$ is not a pole of $G^R$. This means that $G^R$ has a line of poles along the curve $\omega=F(k)$, except at the special points, where the would-be poles are skipped\footnote{The fact that holographic Green's functions have an infinite number of special points was recently shown in~\cite{Grozdanov:2019uhi,Blake:2019otz}.}. 

The precise location of the special points depends on the type of two-point function considered~\cite{Grozdanov:2019uhi,Blake:2019otz,Natsuume:2019xcy}. In particular, for the energy density two-point function, the lowest-lying special point is related to the Lyapunov exponent and butterfly velocity as
\be
\omega_*=i \lambda_L\,,\,\,\,\,k_*=i \frac{\lambda_L}{v_B}\,.
\ee
This seems to be a generic property of holographic systems, being valid even under the presence of higher curvature corrections~\cite{Grozdanov:2018kkt}.

The above discussion is valid for black holes with planar horizons. In those cases, the boundary theory lives in flat space, and we can expand the metric perturbation in terms of plane waves. For black holes with spherical or hyperbolic horizons, we will see that we can expand the metric perturbations in terms of generalized spherical harmonics, with analytically continued angular momentum $L$.\footnote{We thank Richard Davison for pointing this out.} In those cases, 
the pole skipping-point will occur for a special value of $(\omega,L)$ which will also be related to $\lambda_L$ and $v_B$.

In planar black holes, pole-skipping points to a connection between chaos and hydrodynamics. In hyperbolic black holes, it is not even clear if one should expect hydrodynamics behavior. However, our results show that pole-skipping happens even in cases where there is no obvious definition of hydrodynamics, if any.

\subsection{Pole-skipping in hyperbolic black holes}
In this section, we study the pole-skipping phenomenon in (3+1) dimensional Einstein gravity
\be
S=\frac{1}{16 \pi G_N} \int d^4x \sqrt{-g}\left(R+\frac{6}{\ell^2} \right)\,.
\ee
We consider the following hyperbolic black hole solution
\bea
ds^2&=&\frac{1}{z^2}\left(-\mathcal{F}(z)dt^2+\frac{dz^2}{\mathcal{F}(z)} +d\chi^2+\sinh^2\chi d\phi^2\right)\,, \label{eq-hypBH3} \\
\mathcal{F}(z)&=&1-z^2-\frac{1-z_0^2}{z_0^3}z^3\,,
\eea
where $z_0$ denotes the position of the horizon, while the boundary is located at $z=0$. Comparing with \eqref{eq-metricS}, here we use $z=\ell/r$ and use $\mathcal{F}(z)$ to distinguish it from $f(r)$ in (\ref{eq-f}). The Hawking temperature is given by
\be
T=\frac{3-z_0^2}{4\pi \ell z_0}\,.
\label{eq-T2}
\ee
For our purposes, it will be useful to introduce the incoming Eddington-Finkelstein coordinate $v$
\be
v=t-z_*\,,\,\,\,\,\,\,dz_*=\frac{dz}{\mathcal{F}}\,,
\ee
in terms of which the metric becomes
\be
ds^2=-\frac{\mathcal{F}(z)}{z^2}dv^2-\frac{2}{z^2} dv dz+\frac{1}{z^2}\left( d\chi^2+\sinh^2\chi d\phi^2 \right)\,.
\ee

We will be interested in the energy density retarded two-point function $G^R_{T^{00}T^{00}}$ of the corresponding boundary theory. In planar black holes, this quantity is related to fluctuations of the metric field in the sound channel~\cite{Kovtun:2005ev}, which are related to the $vv$ component of Einstein's equations.  In hyperbolic black holes, the decomposition of the metric perturbations into different channels is different from the planar case, but the energy density two-point function is still related to the $vv$ component of Einstein's equations.

We write the metric fluctuations as
\be
\delta g_{\mu \nu}(z,v,\chi,\phi)=\delta \bar{g}_{\mu \nu}(z,\chi,\phi)e^{-i \omega v}\,.
\ee
The pole skipping phenomenon is related to a special property of Einstein's equation near the black hole horizon. More specifically, the constraint imposed by the $vv$ component of Einstein's equations is absent precisely at the special point, leading to the existence of an extra linearly independent incoming solution, that ultimately makes  $G^R_{T^{00}T^{00}}$ infinitely multiple-valued at the special point \cite{Blake:2018leo}. 

To understand how this comes about, we consider a near horizon solution of the form
\be \label{eq-HORIZON}
\begin{split}
\delta \bar{g}_{\mu\nu}(z,\chi,\phi) = \delta g_{\mu\nu}^{(0)}(\chi,\phi)  \,+\,  \delta g_{\mu\nu}^{(1)}(\chi,\phi)(z-z_{0}) \,+\, \mathcal{O}\left[(z-z_{0})^{2}\right] \,.
\end{split}
\ee
The $vv$ component of Einstein's equations reads
\bea
&&2 \left[1+ \frac{\ell}{z_0} \left( 4\pi T-i \omega -\frac{3z_0^{-1}}{\ell}\right) \right] \delta g_{vv}^{(0)}-  \text{csch}^2\chi \, \partial_{\phi}^2 \, \delta g_{vv}^{(0)}- \coth \chi \, \partial_{\chi} \, \delta g_{vv}^{(0)}-\partial_{\chi}^2 \,\delta g_{vv}^{(0)} = \nonumber\\
&&  (2\pi T +i \omega)\left[ 2\coth \chi \, \delta g_{v\chi}^{(0)}+i \omega \left( \text{csch}^2\chi \, \delta g_{\phi \phi}^{(0)}+\delta g_{\chi \chi}^{(0)} \right) + 2 \, \text{csch}^2\chi \, \partial_{\phi} \delta g_{v\phi}^{(0)}+2\, \partial_{\chi}  \delta g_{v\chi}^{(0)}  \right]\nonumber \\
 \label{eq-VVEE}
\eea
For general values of $\omega$, the above equation imposes a constraint involving the horizon values of the metric components $\delta g_{vv}^{(0)}\,, \delta g_{v\chi}^{(0)}\,, \delta g_{\chi\chi}^{(0)}\,, \delta g_{v\phi}^{(0)}$ and $\delta g_{\phi\phi}^{(0)}$. However, when the frequency takes the special value, $\omega =\omega_*:= i 2\pi T$, the metric component $\delta g_{vv}^{(0)}$ decouples from the other components and (\ref{eq-VVEE}) dramatically simplifies
\be
 \left(1-3 z_0^{-2}\right) \delta g_{vv}^{(0)}+ \left( \text{csch}^2\chi \, \partial_{\phi}^2 + \coth \chi \, \partial_{\chi} +\partial_{\chi}^2 \right) \delta g_{vv}^{(0)} =0\,,
\label{eq-constraint}
\ee
taking precisely the same form as  the equation for the shock wave profile (\ref{eq-shockwaveprofile}) for $d=3$ and $\beta = 4\pi \ell z_0/(3-z_0^2)$. Note that $\square_{H_2}=\text{csch}^2\chi \, \partial_{\phi}^2 + \coth \chi \, \partial_{\chi} +\partial_{\chi}^2$.

%\footnote{See, for instance, section V.2 of \cite{Balazs:1986uj}.}

Now, to find the pole skipping point, we write the metric perturbation in terms of {\it generalized} spherical harmonics
\be
\delta g_{vv}^{(0)} (\chi,\phi)= Y_L^M(i \chi, \phi) = e^{i M \phi} P_L^{M}(\cosh \chi)\,,
\ee
where $P_L^{M}$ is an associated Legendre function. Here, we call $Y_L^M(i \chi, \phi)$ {\it generalized} spherical harmonics because the parameter $L$ is unconstrained -- it may even be a complex number, while the index $M$ is an integer. In higher dimensional cases, the above function satisfies the equation
\be
\square_{H_{d-1}} Y_L^M(i \chi, \Omega_{d-2}) = L (L+d-2) Y_L^M(i \chi, \Omega_{d-2})\,.
\ee
With the above ansatz, (\ref{eq-constraint}) becomes
\be
\left[ \left(1-3z_0^{-2} \right)+L(L+1) \right] \delta g_{vv}^{(0)}(\chi,\phi) =0\,.
\label{eq-vv}
\ee
For generic values of $L$, this equation sets the constraint $\delta g_{vv}^{(0)} =0$. However, at the special points
\be
L=L_*^{\pm} :=- \frac{1}{2} \left(1 \pm \sqrt{12z_0^{-2}-3} \right) \,,
\ee
(\ref{eq-constraint}) is identically satisfied, providing no constraint for $\delta g_{vv}^{(0)}$. As explained in~\cite{Blake:2018leo}, the absence of this constraint implies the existence of a second linearly independent incoming solution that ultimately leads to pole-skipping in the energy density two-point correlation functions. 

Interestingly, for generic values of $z_0$, $L_*^{+}$ is related to the butterfly velocity \eqref{eq-vb} in a simple way
\be \label{result123}
L_*^{+}=- \frac{2\pi T }{v_B}\ell\,, \qquad \,v_B=\frac{\sqrt{12z_0^{-2}-3}-1}{4z_0^{-1}}\,,
\ee
Note that $z_0 =  1/\scriptr_\mt{0}$ and $T$ is given by (\ref{eq-T2}). The other solution $L_*^{-}$ is related with $\mu_-$ in \eqref{jhk} and it is irrelevant as discussed below \eqref{jhk}.  For large black holes, i.e., for $z_0 \ll 1$, the special point is related to the butterfly velocity in flat case: $L_*^{+}= -\sqrt{3} z_0^{-1}= - \frac{2\pi T}{v_B} \ell$. 
Note that $L_*^{+}<0$, which is allowed because our ansatz for $\delta g_{vv}^{(0)} (\chi,\phi)$ is proportional to the associated Legendre function, $P_L^M$, for which the parameter $L$ is unconstrained.

\subsubsection*{Higher dimensional cases}
In higher dimensional cases, (\ref{eq-vv}) becomes
\be
\left[L(L+d-2) - \frac{2\pi}{\beta}r_0 (d-1)\right] \delta g_{vv}^{(0)}(\chi,\Omega_{d-2}) =0\,,
\ee
where $r_0=\ell/z_0$. The corresponding pole-skipping point is now 
\be \label{yui}
L = L_*^+:=-\frac{1}{2} \left(d-2 +\sqrt{d\left[ 2(d-1) \frac{r_0^2}{\ell^2}-(d-2) \right]} \right)= -\frac{2\pi T}{v_B} \ell \,.
\ee
with $v_B$ given by \eqref{eq-vb}.\footnote{The other solution $L_*^-:=-\frac{1}{2} \left(d-2 -\sqrt{d\left[ 2(d-1) \frac{r_0^2}{\ell^2}-(d-2) \right]} \right)$ is related with $\mu_-$ in \eqref{jhk} and it is irrelevant as discussed below \eqref{jhk}.}
\subsubsection*{Asymptotic behavior and quasinormal modes}
To understand how the above results are connected with the result for planar black holes, we note that
\be
Y_L^0(i \chi, \phi) \propto P_L(i \chi) \,,
\ee
where $P_L(i \chi)$ is the Legendre function. For large values of $\chi$, we can write
\be
P_L(i \chi) \approx  \cos^L(i \chi) = \left( \cosh \chi \right)^L \approx e^{L \chi}\,.
\ee
This shows that, for large values of $\chi$, the metric perturbations behave as 
\begin{equation}
\sim e^{-i\omega v +L\chi}  = e^{-i\omega v + i \frac{L}{i\ell} (\ell \chi)} \ \rightarrow  \  e^{-i\omega v + i k (\ell \chi)} \,,
\end{equation}
where we identified  $-i L/\ell \equiv k $. In terms of $k$, the pole-skipping point is given by $k=k_*:=i\frac{2\pi T}{v_B}$, which is  the same from as the flat case. Note that from this relation we may also identify the butterfly velocity $v_B$ from $L$, i.e. $v_B = -2\pi T\ell/L$, which is consistent with \eqref{yui}. 

Moreover, at the pole-skipping, we recover the shock wave transverse profile, i.e., $e^{L_* \chi} = e^{-\frac{2\pi T}{v_B}\ell\chi}$. Based on the observed parallelism between the metric perturbation $\delta g_{vv}^{(0)}$ and the shock wave transverse profile~\cite{Blake:2018leo} in the case of planar black holes, it seems that $L_*$ has to be negative for the metric perturbations to have the correct asymptotic behavior. i.e. the negative value of $L_*$ leads to metric perturbations that decay exponentially when we move away from the source.

\begin{figure}[]
 \centering
     {\includegraphics[width=7.3cm]{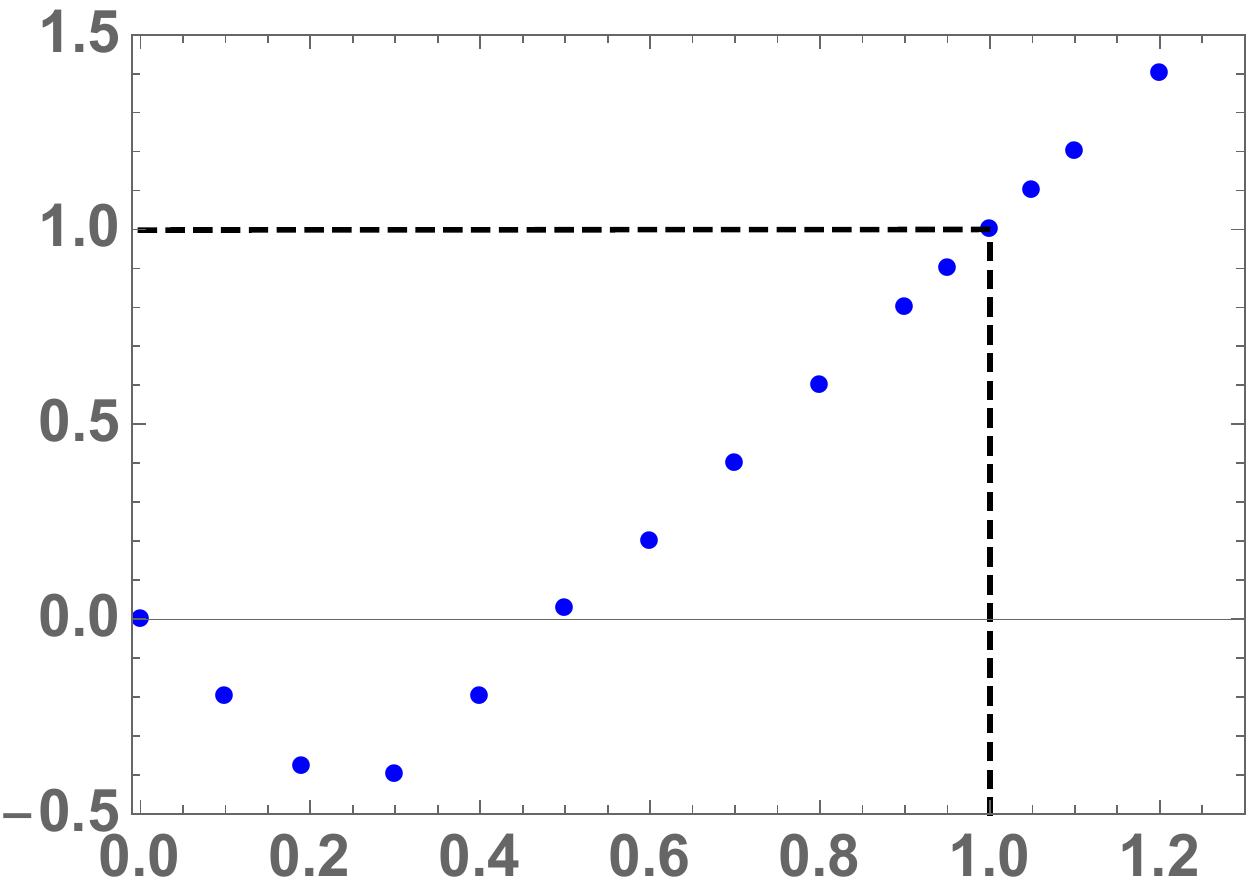} } 
           \put(-110,-17){\large $\frac{v_B\text{Im}(k)}{2\pi T}$}
           \put(-235,74){\large $\frac{\text{Im}(\omega)}{2\pi T}$}
          \caption{Sound channel quasinormal modes of the hyperbolic black hole defined in (\ref{eq-hypBH3}). The blue dots represent the zeros of $a(\omega,k)$ in \eqref{qnm1}. They form a  line that passes though the special point $(\omega_{\star}, k_{\star}) =i (2 \pi T, 2 \pi T / v_{B})$ with $v_{B}$ given in \eqref{result123}. Here, $k :=-i L/\ell$, and we set $z_{0}=1$. For simplicity, we used the asymptotic form of the Legendre functions in the ansatz for the metric perturbations.} \label{fig-QNMF}
\end{figure}

Finally, by considering angular independent perturbations, and taking the large $\chi$ limit of the equations of motion, we can define the sound channel, just like in the planar case. This channel involves $\delta g_{vv}\,,\delta g_{vz}\,, \delta g_{zz}\,, \delta g_{v\chi}\,,\delta g_{z\chi}\,, \delta g_{\chi \chi}\,, \delta g_{\phi \phi}\,$. As another independent crosscheck, we numerically  computed the quasinormal modes of this `emergent' sound channel and confirmed that the line of poles of $G^R_{T^{00}T^{00}}$ precisely passes through the  pole-skipping point $(\omega_*,k_*)$, where $k_* = -iL_*/\ell$. It confirms again our result, \eqref{eq-vb} or \eqref{result123}. See figure \ref{fig-QNMF}.

\section{Discussion } \label{sec-disc}

In this paper, we have studied the scrambling properties of $(d+1)-$dimensional hyperbolic black holes. We gave a precise derivation of OTOCs for a Rindler-AdS$_{d+1}$ geometry, which is dual to a $d-$dimensional CFT in hyperbolic space with inverse temperature $\beta=2\pi \ell$. We found 

\be
\text{OTOC}(t,b)=1+\frac{ i\,c_1 \Delta_{W} \Delta_V}{ \epsilon_{13} \epsilon^\mt{*}_{24}}e^{ \frac{1}{\ell}(t-t_*)-\frac{(d-1)}{\ell}b}\,,\,\,\,\,\,b := \ell d({\bf x_1,x_2})\,,
\ee
which implies 
\be
\lambda_L=\frac{1}{\ell} = 2\pi T  \,,\,\,\,\,\, v_B=\frac{1}{d-1}\,.
\ee
The above result perfectly matches the corresponding CFT results~\cite{Perlmutter:2016pkf}. 

For more general hyperbolic black holes, we calculated the phase shift, which essentially controls the form of the OTOCs, and from which we can extract the Lyapunov exponent and butterfly velocity. We found
\be
\lambda_L = 2 \pi T\,,\,\,\,\,\,\, v_B(\scriptr_\mt{0})  = \frac{\sqrt{d \,\big[ 2\scriptr_\mt{0}^2 (d-1)-(d-2)\big] }-(d-2)}{2(d-1) \scriptr_\mt{0}}\,,\,\,\,\,\,      \scriptr_\mt{0}=\frac{r_0}{\ell}\,.
\ee
In section \ref{sec-poleskip}, we checked that the above result can also be obtained from a pole-skipping analysis in two ways: i) the analytic near horizon condition, ii) the numerical quasinormal mode computation. Contrary to the flat case, we expanded the metric perturbation in terms of spherical harmonics, with analytically continued angular momentum  $L$, instead of plane waves. It is interesting that the pole-skipping analysis reveals a connection between chaos and hydrodynamics also in hyperbolic black holes even though it is   not clear if one should expect hydrodynamics behavior in hyperbolic space.

In our pole-skipping analysis, we consider metric perturbations which coupled to $\delta g_{vv}$. These perturbations form a sector that is analogous to the sound channel of planar black holes. In the case of planar black holes, pole-skipping happens not only in the sound channel, but also in the other channels \cite{Grozdanov:2019uhi,Blake:2019otz,Natsuume:2019xcy}. It would be interesting to check if this also happens for other channels of hyperbolic black holes.

The temperature dependence of the butterfly velocity is shown in figure \ref{fig-vb1}. The butterfly velocity is zero at $T=0$, and increases as  $T$ increases, quickly approaching the asymptotic value $\sqrt{\frac{d}{2(d-1)}}$. This asymptotic value precisely coincides with the butterfly velocity for a planar Schwarzschild black hole in $(d+1)$ dimensions~\cite{BHchaos3}. This is expected, because the very large temperatures occur for very large black holes $\frac{r_0}{\ell} \gg 1$ (see (\ref{eq-T})), for which the geometry of the horizon should be approximately flat. Moreover, for $T=\frac{1}{2\pi \ell}$, we recover the Rindler-AdS$_{d+1}$ result: $v_B=\frac{1}{d-1}$.

In the context of Einstein gravity, the butterfly velocity was shown to be bounded for isotropic planar black holes satisfying the Null Energy Condition, with the bound given by the Schwarzschild result~\cite{Mezei:2016zxg}
\be
v_B \leq \sqrt{\frac{d}{2(d-1)}}\,.
\ee
Our result suggests that this bound might also be valid for black holes with non-planar horizons\footnote{This bound was shown to be violated by anisotropy~\cite{Jahnke:2017iwi,Avila:2018sqf,Fischler:2018kwt,Baggioli:2018afg} and higher curvature corrections~\cite{Grozdanov:2018kkt}. For other interesting effects of higher curvature corrections on $v_B$, see, for instance~\cite{Alishahiha:2016cjk}.}. It would be interesting to check whether this bound can be derived in these cases.

\section*{Acknowledgments}

It is a pleasure to thank Wyatt Austin for useful discussions and M\'{a}rk Mezei, Richard Davison, and Sa\v{s}o Grozdanov for useful correspondence. This work was supported in part by Basic Science Research Program through the National Research Foundation of Korea(NRF) funded by the Ministry of Science, ICT \& Future Planning(NRF2017R1A2B4004810) and GIST Research Institute(GRI) grant funded by the GIST in 2019.

\appendix

\section{Full solution for the shock wave transverse profile}\label{AppA}

The equation of motion for the shock wave transverse profile reads
\be
\left[ \square_{H_{d-1}}-\frac{2\pi}{\beta} r_0 (d-1) \right] h({\bf x})=-\frac{8 \pi G_{N}}{r_0^{d-3}} p^\mt{V}\delta({\bf x,0})\,.
\label{eq-shockwaveprofile2}
\ee
Assuming that $h$ does not depend on the angular coordinates $\Omega_{d-2}$ and taking ${\bf x}\neq 0$, this equation becomes\footnote{Here we use that $\square_{H_{d-1}}= \partial_{\chi}^2+(d-2) \coth \chi \partial_{\chi}+\frac{1}{\sinh^2\chi} \square_{S^{d-2}}$.}
\be
\left[ \partial_{\chi}^2+(d-2) \coth \chi \partial_{\chi}-\frac{2\pi}{\beta}r_0 (d-1) \right] h(\chi)=0\,.
\ee
This equation has an exact solution
\bea
h(\chi)=&& i^{1-d} \left(-\text{sech}^2(\chi)\right)^{\frac{1}{4} (a-2)} \tanh ^{\frac{1}{2}-\frac{d}{2}}(\chi) \tanh ^2(\chi)^{\frac{1}{4} (-d-1)} \text{sech}^2(\chi)^{d/4} \times \nonumber \\
&&\Bigg[ c_2 \tanh ^2(\chi)^{3/2} \, _2F_1\left(\frac{a-d+4}{4} ,\frac{a-d+6}{4} ;\frac{5-d}{2};\tanh ^2(\chi)\right)+\nonumber \\
&&\,\,\,\,\,\,c_1\,  i^{d+1} \tanh ^2(\chi)^{d/2} \, _2F_1\left(\frac{a+d-2}{4} ,\frac{a+d}{4} ;\frac{d-1}{2};\tanh ^2(\chi)\right) \Bigg] \,,
\label{eq-fullsolution}
\eea
where 
\be
a := \sqrt{ (d-2)^2+\frac{8\pi}{\beta}r_0 (d-1)}\,,
\ee
and $c_1$ and $c_2$ are arbitrary constants.
The asymptotic solution for large values of $\chi$ can then be obtained as 
\be
h(\chi) \sim \, e^{-\mu_+  \chi} \,, \quad   e^{-\mu_-  \chi}\,,
\ee
where 
\begin{equation} \label{jhk}
\mu_{\pm} := \frac{1}{2}\left(d-2 \pm \sqrt{(d-2)^2+\frac{8\pi r_0}{\beta}(d-1)} \right)\,, \qquad \mu := \mu_+ \,.
\end{equation}
Here, we discard the second solution because $\mu_- < 0$ always, which means the perturbation grows when we move away from the source, instead of decreasing. For notational simplicity, we define $\mu := \mu_+$.

As a consequence of the $SO(d-1,1)$ symmetry of $H_{d-1}$, the shock wave transverse profile only depends on the geodesic distance $\chi=d({\bf x,0})$ between ${\bf x}$ and the position of the source, which take as ${\bf 0}$ in (\ref{eq-shockwaveprofile2}). If we write the right hand side of (\ref{eq-shockwaveprofile2}) with a source proportional to $\delta({\bf x,x'})$, the $SO(d-1,1)$ isometry of hyperbolic space allows us to conclude that $h(d({\bf x,x'})) = \text{constant} \, \times \, e^{-\mu \, d({\bf x,x'})}$.

\bibliographystyle{JHEP}
%\bibliography{hybBH}

\providecommand{\href}[2]{#2}\begingroup\raggedright\endgroup

\end{document}